\documentclass[preprint, authoryear,%
]{elsarticle}

\usepackage{amssymb}

\usepackage{amsmath} 
\usepackage{graphicx} 

\usepackage{xargs} 
\usepackage{xcolor} 

\usepackage{bm}

\ifdoubleblind
 \newcommand{\ficosManuscript}{Authors anonymous, manuscript}
\else
 \newcommand{\ficosManuscript}{Lignell et al., manuscript}
\fi

\usepackage{subcaption}
\usepackage{rotating}
\usepackage{algorithm}
\usepackage{algpseudocode}

\DeclareMathOperator*{\argmax}{arg\,max}

\begin{document}

%

%

\title{Bayesian Calibration and Uncertainty Quantification for a Large Nutrient Load Impact Model}

\author[1]{Karel Kaurila\corref{cor1}}
\ead{karel.kaurila@helsinki.com}

\author[2]{Risto Lignell}
\ead{risto.lignell@gmail.com}

\author[3]{Frede Thingstad}
\ead{Frede.Thingstad@uib.no}

\author[2]{Harri Kuosa}
\ead{Harri.Kuosa@syke.fi}

\author[1,4]{Jarno Vanhatalo}
\ead{jarno.vanhatalo@helsinki.fi}

\cortext[cor1]{Corresponding author}

\affiliation[1]{organization={%
	Department of Mathematics and Statistics,
	Faculty of Science,
	University of Helsinki},
	addressline={PL 68, Pietari Kalmin katu 5},
	postcode={00014 Helsingin yliopisto},
	city={Helsinki},
	country={Finland}}

\affiliation[2]{organization={%
 Finnish Environment Institute},
	addressline={Latokartanonkaari 11},
	postcode={00790},
	city={Helsinki},
	country={Finland}}
	
\affiliation[3]{organization={%
Department of Biological Sciences,%
University of Bergen},
	addressline={Postboks 7803},
	postcode={5020},
	city={Bergen},
	country={Norway}}

\affiliation[4]{organization={%
	Organismal and Evolutionary Biology Research Programme,
	Faculty of Biological and Environmental Sciences,
	University of Helsinki},
	addressline={P.O. Box 65}, 
	postcode={00014 University of Helsinki}, 
	city={Helsinki},
	country={Finland}}

\begin{abstract}
Nutrient load simulators are large, deterministic, models that simulate the hydrodynamics and biogeochemical processes in aquatic ecosystems.
They are central tools for planning cost efficient actions to fight eutrophication since they allow scenario predictions on impacts of nutrient load reductions to, e.g, harmful algal biomass growth. 
Due to being computationally heavy, the uncertainties related to these  predictions are typically not rigorously assessed though. 
In this work, we developed a novel Bayesian computational approach for estimating the uncertainties in predictions of the Finnish coastal nutrient load model FICOS. 
First, we constructed a likelihood function for the multivariate spatiotemporal outputs of the FICOS model.
Then, we used Bayes optimization to locate the posterior mode for the model parameters conditional on long term monitoring data. 
After that, we constructed a space filling design for FICOS model runs around the posterior mode and used it to train a Gaussian process emulator for the (log) posterior density of the model parameters. 
We then integrated over this (approximate) parameter posterior to produce probabilistic predictions for algal biomass and chlorophyll $a$ concentration under alternative nutrient load reduction scenarios. 
Our computational algorithm allowed for fast posterior inference and the Gaussian process emulator had good predictive accuracy within the highest posterior probability mass region.
The posterior predictive scenarios showed that the probability to reach the EU’s Water Framework Directive objectives in the Finnish Archipelago Sea is generally low even under large load reductions. 
\end{abstract}


\maketitle

\section{Introduction}

Human induced excess nutrient loads significantly affect aquatic ecosystems worldwide.
They lead to eutrophication, which occurs when nutrient enrichment increases plankton standing stocks, including biomass of harmful algae \citep{Kiirikki+etal:2001,Murray+etal:2019}.
For this reason, many countries aim to attain a good ecological status in aquatic ecosystems via nutrient load reductions. 
Implementing these actions is challenging since they require expensive and large scale national measures, such as regulations to agricultural and aquacultural practices.
Hence, the expected effects of alternative actions are commonly analysed and compared with nutrient load impact simulators \citep{Kiirikki+etal:2001,Murray+etal:2019}. 
These are deterministic models that simulate the hydrodynamics and biogeochemical processes in an ecosystem, and can be used to predict the development of algal biomass under alternative nutrient load scenarios. 
Because these models aim to simulate large scale real world systems, such as the the Archiplego Sea in the Baltic Sea in this work (Figure~\ref{fig:FICOS}), they are typically computationally extremely heavy. 
Due to the large computational time needed to produce the outputs from these models, they pose challenge for model calibration and uncertainty estimation.
As a result, the uncertainties related to these models' predictions are typically not rigorously assessed. 

If the uncertainties in nutrient load impact simulator predictions are assessed at all, a typical approach is to use  Generalized Likelihood Uncertainty Estimation (GLUE) methodology \citep{bevenEquifinalityDataAssimilation2001} for calibrating the simulator. 
GLUE is an \textit{informally Bayesian} approach that aims to circumvent the diffult problem of formally defining a likelihood for observations used to calibrate a complex hydrological model \citep{bevenGLUE20Years2014}. 
It has, however, been criticised for being incoherent with Bayesian inference \citep{mantovanHydrologicalForecastingUncertainty2006, stedingerAppraisalGeneralizedLikelihood2008} for which reason it leads to inconsistent uncertainty estimates as well.
Another common approach is to do sensitivity analysis for the simulator outputs without formal uncertainty quantification and proper probabilistic interpretation of their predictions. 
For example, \citet{taylorModellingImpactsAgricultural2016} modeled the impact of agricultural management on water quality in this manner. 
A few more formally Bayesian methods for uncertainty quantification in the hydrology literature have been proposed as well. 
\citet{thiemannBayesianRecursiveParameter2001} used an initially defined grid of candidate parametrisations for approximating the posterior for model parameters.
Their approach was, however, limited to computationally non-expensive simulators, small number of parameters, and a situation where the limits of the highest posterior probability mass region are well known \emph{a priori}.
\citet{Vanhatalo+etal:2013} assessed the spatiotemporal biases in a large scale nutrient load impact model using a non-parametric Bayesian approach, and estimated the induced uncertainties in the impacts of alternative management actions in the Gulf of Finland context. 
Despite using the Bayesian approach, their treatment ignored the model parameter uncertainties and quantified the uncertainties through model discrepancy terms only. 

In this work, we make a step further from these existing approaches by developing a novel Bayesian inference algorithm to calibrate the key unknown biological parameters of a large scale Finnish coastal nutrient load model (FICOS; \ficosManuscript) and to estimate the uncertainties in the FICOS model predictions induced by the uncertainties in the model parameters. 
The FICOS model is developed for the Archipelago Sea in the Northern Baltic Sea to simulate the nutrient loads and the resulting eutrophication responses in that coastal system. 
It includes dynamic biogeochemical model (BGC), which is coupled with a high-resolution 3D hydrodynamic coastal model \citep{Tuomi+etal:2018}.
The FICOS model is currently in operational use in the Finnish Environment Institute (Syke) in management and policy planning. 
However, the current parameterization of the model does not follow a statistically rigorous approach, though the key parameters are based on theoretical, empirical and statistical examinations \citep{Lignell+etal:2013}. 
Moreover, the inherent uncertainties in the model formulations, parameterisations, and predictions are not accounted for in these analyses.

In our method, we first defined prior distributions for the calibration parameters and constructed likelihood function for them by linking the multivariate spatiotemporal outputs of FICOS with long term monitoring data.
After this, we developed an Bayes optimization (BO) algorithm \citep{jones1998efficient} to locate the maximum a posterior estimate of these BGC parameters, and construct a Gaussian process emulator \citep{Kennedy+OHagan:2001} to approximate the (log) posterior distribution around the mode. 
For last, we marginalized over this (approximate) posterior with Markov chain Monte Carlo and importance sampling to produce probabilistic predictions for algal biomass and chlorophyll~$a$ concentration under alternative nutrient load reduction scenarios. 
Since the calibration parameters govern the algal growth dynamics, accounting for the uncertainty in them provides means to assess the uncertainty in nutrient load reduction scenarios.

Our approach is similar to BO guided approximate Bayesian computation (ABC) and likelihood free inference (LFI), methods originally developed to infer parameters of complex stochastic simulators for which the likelihood function is not analytically available \citep[e.g.,][]{holden2018abc, jarvenpaa2020batch,tianGaussianProcessEmulators2017, gutmann2016bayesian}. 
Previously, special versions of ABC have also been used to build surrogates for exact likelihood functions in the hydrology literature \citep[e.g.,][]{vrugtDiagnosticModelCalibration2013, sadeghApproximateBayesianComputation2014}. In this respect, the GLUE method that is commonly used for estimating uncertainties in hydrology can also be seen as a special case of ABC \citep{nottGeneralizedLikelihoodUncertainty2012}. 
However, the ABC approaches for hydrological models have not used BO to optimize the parameters. 
Similarly to BO guided ABC approches in other applications \citep[e.g,][]{doBayesianOptimizationassistedApproximate2023}, we use BO to guide the inference to the highest posterior mass region and then use Gaussian processes to emulate the posterior probability density around the mode.
However, unlike in ABC algorithms we formulate an analytic likelihood function for our model  parameters, which leads to exact posterior distribution. 
Gaussian process emulators aided inference in our approach is also similar to some earlier works on uncertainty quantification of complex, multioutput and dynamic, computer models \citep[e.g.,][]{Kennedy+OHagan:2001,Conti+Ohagan:2010,Salter+etal:2019,Coveney+etal:2020}. 
However, instead of predicting the output of computationally heavy computer model, as done in these examples, our Gaussian process emulator represents the log posterior density of its parameters.

Our approach is also the first serious attempt to quantify the uncertainty in nutrient load impact models for the Baltic Sea -- one of the most eutrophicated sea areas in the world.
As such, the application of our approach is novel and of utmost importance for the sustainable management of the Baltic Sea since, for the first time, we are able to produce scientifically rigorous quantitative estimates for the effects of alternative nutrient load reductions on eutrophication in the Baltic Sea.

\section{The nutrient load impact model and the calibration data}

\subsection{FICOS nutrient load model}

\begin{figure}[t]
\vspace{.3in}
\centerline{
	\includegraphics[width=5in]{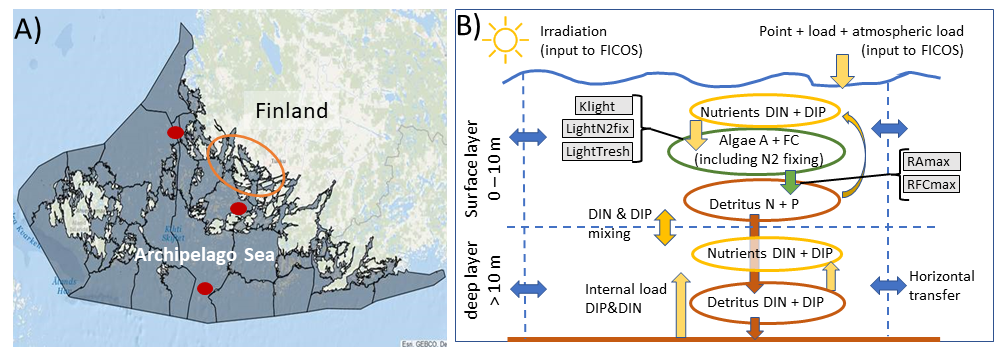}
	}
\caption{(A) The spatial domain of the FICOS model with the sub-basin polygons used in this work (grey tiles), the approximate area of the inner Archipelago Sea used in the scenario analyses (inside yellow oval), and, from north to south the Brändö, Seili and Utö intensive stations (red dots). (B) The overview of the key processes described by the biogeochemical model compartment of the FICOS model and the position of the calibration parameters (grey boxes) in it.}\label{fig:FICOS}
\end{figure}

The newly developed FICOS model system (\ficosManuscript) is a deterministic computer model that simulates the biogeochemical processes in the Archipelago sea area in the Baltic Sea (Figure~\ref{fig:FICOS}). 
The full model includes sub-models for the nutrient and freshwater load from drainage basin \citep[WSFS-VEMALA model,][]{Korppoo+etal:2017}, 3-dimensional hydrodynamics \citep[][]{Tuomi+etal:2018} including nutrient transport and background load, model for biogeochemistry \citep[BGC, modified from][]{Kiirikki+etal:2001} and an empirical-statistical model for the description of internal nutrient load from sediments. 
FICOS operates with a time step of one day
and its spatial resolution can be altered from 0.25 nm grid up to polygons of sub-basin scale water formations (from tens to hundreds of square nautical miles, shown in Figure~\ref{fig:FICOS}).
The water formations consist of two vertical layers: $10$ m deep surface layer and deep layer that contains everything below the surface layer. 
Due to computational constraints we use the coarsest sub-basin level resolution in this work.

The FICOS simulator is programmed in Fortran and uses the HDF5 file format \citep[see][for overview]{folkOverviewHDF5Technology2011} for its inputs and outputs. 
In the coarsest resolution used for calibration in this work, the nutrient loading and boundary condition files have a combined size of 132 MB while the outputted prediction file has size 43 MB. 
Running FICOS for the calibration time period (2006-2015, Section~\ref{sec:calibration_data}) for a single parametrization had a 4 minute run time on our computer with Intel(R) Xeon(R) Gold 6240R CPUs at 2.40GHz clock speed.
To illustrate the computational constraints with finer resolutions, running one simulation on a setting where approximately 1/3 of the study region was modelled in the finer 0.25 nm resolution took our computer over 24 hours and the resulting output had size 6.6 GB.

The FICOS model system can be used to realistically simulate the annual development of algal biomass, chlorophyll~$a$ and phosphorus and nitrogen concentrations (inorganic and total), and the impact of changes in external nutrient load on these variables (Figure~\ref{fig:FICOS}B).
The nutrient load, water flow at the boundaries, and solar radiation are considered as inputs to FICOS. 
As outputs, the model estimates the nutrient concentrations in deep and surface water layers as well as the biomass of N$_2$ fixing cyanobacteria and other algae in the photic surface layer.
For this work, we identified five key unknown parameters of the BGC model component for calibration through formal (approximate) Bayesian inference.
These parameters are related to algal death rate (RAmax, RFCmax), light limitation of algal production (Klight), and cyanobacterial N$_2$ fixation (N2fixThres, ProdThres).

\subsection{Prior distributions for calibration parameters}

We gave each of the calibration parameters a weakly informative log Gaussian prior based on an expert assessment directed by earlier literature and experimental results on nutrient dynamics in the Baltic Sea. 
The expert assessment contained the most likely parameter value and an estimate for the upper bound. 
We then coded these, respectively, as the prior mean and the 95\% upper credible bound of a parameter. 
The resulting mean, $\mu$, and standard deviation, $\sigma$, of these log Gaussian prior distributions were  $\mu = \ln 10, \sigma = \frac{\ln 2}{2}$ for Klight, $\mu = \ln 15, \sigma = \frac{\ln 2}{2}$ for N2fixThres, $\mu = \ln 10, \sigma = \frac{\ln 2}{2}$ for ProdThres, $\mu = -2, \sigma = 0.4$ for RAmax, and $\mu = -2, \sigma = 0.4$ for RFCmax.

\subsection{Calibration data}\label{sec:calibration_data}

The posterior inference for the key parameters of the BGC model was done based on long term monitoring observations on dissolved inorganic nitrogen (DIN) and phosphorus (DIP), Chla, and phytoplankton species (filamenthous cyanobacteria and other algae). 
These data were obtained from (\ficosManuscript) and were mainly based on the Finnish Environment Institute (Syke) monitoring database.
Key intensive annual monitoring data for the BGC model calibration were data from three stations in Northern, middle and Southern Archipelago Sea: Brändö, Seili, and Utö (Figure~\ref{fig:FICOS}A). 
All data were collected between years 2006 and 2015, and for the nutrients, DIN and DIP, the data were collected from two layers: surface layer, which is at most $10$~m from the surface, and deep layers, which are at least $20$~m from the surface.

\section{Calibration model}\label{sec:calibModel}

\subsection{Likelihood for non-censored observations}\label{sec:noncensoredLikelihood}

All the calibration parameters were constrained to be positive for biological realism, so we denote the log of them by $\boldsymbol\theta\in \mathbb{R}^d$,
where $d=5$ in our case. 
Because the FICOS outputs and observations are strictly non-negative and the variance of observations around their expectation increases with increasing mean,
we modelled the prediction residuals to be log normally distributed. 
Hence, we denote the FICOS model (log) output by $f: \mathbb{R}^d \mapsto \mathbb{R}^{2\times D\times T\times N}$, where $D=5$ is the number of FICOS output variables (DIN, DIP, chlorophyll $a$, and two algae species concentrations; Section~\ref{sec:calibration_data}) for two depth layers (deep and surface) as well as for $T$ days and $N$ spatial polygons/cells. The time span of the FICOS model output, $T$, corresponds to the calibration time period from 2006 to 2015. 
Similarly, we denote by $\mathbf{y}\in \mathbb{R}^n$ the (log) calibration data and by $\mathbf{j}$, $\mathbf{k}$, $\mathbf{t}$, $\mathbf{r}$ the vectors of indices such that the $i$'th observation, $y_i$, corresponds to the FICOS output $f(\boldsymbol\theta)_{\mathbf{j}(i), \mathbf{k}(i), \mathbf{t}(i), \mathbf{r}(i)}$, where $\mathbf{j}(i)$ is the depth layer where the observation was taken, $\mathbf{k}(i)$ is the observed variable (e.g., DIN concentration), $\mathbf{t}(i)$ is the day of the observation, and $\mathbf{r}(i)$ is the polygon where the $i$th observation was made.

We treated the observations as conditionally independent given the FICOS output and modeled them with a log Gaussian distribution 
\begin{equation}\label{eq:obs_model}
 \ln y_{i} \sim \mathcal{N}(\ln f(\bm{\theta})_{\mathbf{j}(i), \mathbf{k}(i), \mathbf{t}(i), \mathbf{r}(i)},\sigma^2_{\mathbf{j}(i), \mathbf{k}(i)}).
\end{equation}
The variance parameter of the observation model accounts for both measurement errors and the natural variation around the mean predicted by the FICOS model.
Because the measurement accuracy for observations depends on the type of observation and the depth from where the measurement was taken we used a separate error variance $\sigma^2_{j,k}$ for each vertical layer $j$ and variable $k$. 

After standardizing each variable, we then gave their variance parameters a scaled inverse chi-squared prior
\begin{equation}\label{eq:sigma_prior}
 \sigma^2_{j,k} \sim \mathrm{inv}-\chi^2(\nu_0=4, \tau_0^2=1)
\end{equation}

When we marginalize over $\sigma^2_{j,k}$, the distribution for a vector of all observations from the variable $k$ at the depth layer $j$, denoted by $\mathbf{y}_{j,k}$, is the multivariate $t$ distribution \citep{kotzMultivariateDistributionsTheir2004} so that $\mathbf{y}_{j,k}|\bm\theta \sim \mathbf{t}_n(\mathbf{f}(\boldsymbol\theta)_{j,k}, \bm\Sigma, \nu)$, 
where the location parameter $\mathbf{f}(\boldsymbol\theta)_{j,k}$ is the vector of the FICOS model outputs corresponding to the observations $\mathbf{y}_{j,k}$, the scale matrix $\bm\Sigma$ and $\nu$ is the degrees of freedom.
The scale matrix $\bm\Sigma = \tau_{j,k}^2 \mathbf{R}$ is defined though a scale parameter $\tau_{j,k}$ and a correlation matrix $\mathbf{R}$.
Since we standardized the observations in $\mathbf{y}_{j,k}$ (and did correspondingly to $\mathbf{f}_{j,k}$), we used $\tau_{j,k}^2 = \tau_0^2 = 1$ for all $j,k$ and a priori $\bm\Sigma = \mathbf{R}=\mathbf{I}_n$.

For the phytoplankton concentrations, all observations were non-censored. Thus, for the variables FC (filamentous cyanobacteria) and A (other algae), the observation model is simply the multivariate $t$ density, giving us the log likelihood function:
\begin{equation}\label{eq:lik_non-censored}
 \mathcal{L}_{j,k}(\boldsymbol\theta|\mathbf{y}) = -\frac{\nu_0+n_{j,k}}{2}
 \ln\left(1+\frac{(\mathbf{y}_{j,k}-\mathbf{f}_{j,k}(\bm\theta))^T(\mathbf{y}_{j,k}-\mathbf{f}_{j,k}(\bm\theta))}{\nu_0\tau_0^2}\right).
\end{equation}

\subsection{Likelihood for censored and non-censored observations}\label{sec:lik_censored}

The measurement sensitivity for the nutrient mass concentrations, DIN and DIP, was such that quantities below thresholds $a_\textrm{DIN} = 10$~$\mu$ g N/l and $a_\textrm{DIP} = 3$~$\mu$ g P/l, respectively, could not be reliably measured. %
Hence, we treated all DIN and DIP observations below these thresholds as censored, meaning that we only knew that the measured quantity is below the censure boundary $a$, but we did not know the actual value.
We included censored observations in our likelihood through their cumulative density functions as follows.

We denote the vectors of true nutrient mass concentrations corresponding to the censored observations by $\mathbf{f}_\text{c}$ and the non-censored observations by $\mathbf{f}_\text{p}$. Similarly, we denote the unobserved values of the censored measurements by $\mathbf{y}_\text{c}$ and the (observed) values of the non-censored measurements by $\mathbf{y}_\text{p}$. Their joint distribution conditional on $\bm\theta$ is then
\begin{equation}
\begin{bmatrix} \mathbf{y}_\text{p} \\ \mathbf{y}_\text{c} \end{bmatrix} 
| \bm\theta \sim \mathbf{t}_{n} \left(
\begin{bmatrix} \mathbf{f}_\text{p}(\bm\theta) \\ \mathbf{f}_\text{c} (\bm\theta) \end{bmatrix},
\tau_0^2 \mathbf{I}_n, \nu_0 \right)
\end{equation}
where $n = n_\text{p} + n_\text{c}$ is the number of all observations. 
This joint distribution can be partitioned into a marginal distribution of non-censored measurements and a conditional distribution of the unobserved measurements of the censored observations given the non-censored measurements
$p(\mathbf{y}_\text{c},\mathbf{y}_\text{p}|\bm\theta) =  p(\mathbf{y}_\text{c}|\mathbf{y}_\text{p},\bm\theta)p(\mathbf{y}_\text{p}|\bm\theta)$ where 
\begin{align}
\mathbf{y}_\text{p} | \bm\theta  &\sim \mathbf{t}_{n_\text{p}} \left(
\mathbf{f}_\text{p}(\bm\theta) , \tau_0^2 \mathbf{I}_{n_\text{p}} , \nu_0 \right) \nonumber \\
\mathbf{y}_\text{c} | \bm\theta, \mathbf{y}_\text{p} & \sim \mathbf{t}_{n_\text{c}} \left( \mathbf{f}_\text{c}(\bm\theta) , \tau^2_{\text{c} |\text{p}} \mathbf{I}_{n_\text{c}} , \nu_{\text{c} |\text{p}} = \nu_0 + n_\text{p} \right) \label{eq:censored_conditional_mvt}
\end{align}
and $\tau_{\text{c} |\text{p}} = \sqrt{\frac{\nu_0\tau_0^2+ (\mathbf{y}_p - \mathbf{f}_p)^T (\mathbf{y}_p - \mathbf{f}_p) }{\nu_0+n_p}}$ \citep{genzComputationMultivariateNormal2009,dingConditionalDistributionMultivariate2016} . 
The censored observations are of the form $\mathbf{y}_{\text{c}} \leq \mathbf{a}$ for which the conditional probability, given $\mathbf{y}_p$, can be written as \mbox{\citep[][see \ref{sec:appendix_censored_loglik} for details]{genzComputationMultivariateNormal2009}}:
\begin{align} 
\Pr( \mathbf{y}_\text{c} \leq \mathbf{a} | \bm\theta, \mathbf{y}_\text{p} ) & = \Pr(y_{\text{c},i} \leq a_i, \dots, y_{\text{c},n} \leq a_n | \bm\theta, \mathbf{y}_\text{p}) \label{eq:lik_censored} \\ 
&= \int_0^1 \prod_{i=1}^n \Phi \left(\frac{\chi^{-1}_{\nu_{\text{c} |\text{p}} }(t)}{\tau_{\text{c} |\text{p}}\sqrt{\nu_{\text{c} |\text{p}} }} \left(a_i -f_i(\bm\theta) \right) \right) dt \nonumber
\end{align}
where $\chi_\nu^{-1}$ is the inverse CDF of the $\chi$ distribution with $\nu$ degrees of freedom.

%
When we add this to the log probability density for non-censored observations~\eqref{eq:lik_non-censored}, we get the full log-likelihood:
\begin{align}
 \mathcal{L}(\boldsymbol\theta|\mathbf{y}_p, \mathbf{y}_c) = -\frac{\nu_0+n_\text{p}}{2}
 \ln\left(
 1+\frac{\bm\epsilon_\text{p}(\bm\theta)^T \bm\epsilon_\text{p}(\bm\theta)}{\tau_0^2\nu_0}
 \right) \nonumber \\+
 \ln \int_0^1 \prod_{i=1}^{n_\text{c}} \Phi \left(\frac{\chi^{-1}_{\nu_{\text{c} |\text{p}}}(t)}{\tau_{\text{c} |\text{p}}\sqrt{\nu_{\text{c} |\text{p}} }} \left(a_i -f_i(\bm\theta) \right) \right) dt, \label{eq:loglik_full}
\end{align}
where $\bm\epsilon_\text{p}(\bm\theta) = \mathbf{y}_\text{p}-\mathbf{f}_\text{p}(\bm\theta)$.

\section{(Approximate) Bayesian Inference}

\subsection{Bayesian optimization (BO) to locate the posterior mode}\label{sec:BO}

\subsubsection{Summary of the algorithm}\label{sec:BOsummary}

We used a BO algorithm to search the mode of the posterior distribution of the calibration parameters.
In this subsection, we summarize the key steps in our algorithm while explaining 
the technical implementation in more detail in the subsequent subsections.

We used BO to minimize the objective function
\begin{equation}\label{eq:objective}
g(\bm\theta) = h\left( -\ln p(\bm\theta) - \mathcal{L}(\bm\theta|\mathbf{y}) \right)
\end{equation}
where $h(\cdot)$ was either the log transformation, i.e. $h(\cdot) = \log(\cdot)$, in the early phase of the optimization, or the identity function in the last phase of the optimization (to be explained in more detail below).
We started the optimization with the log-minus-log posterior since within the initially large search limits the log posterior density varied so much that we could not get good fit to it with our Gaussian process (GP) surrogate models (i.e., emulator).
In BO, the search of a function's minimum is aided by modelling the objective function as a stochastic process, most commonly GP \citep{jones1998efficient}. 
Modeling the objective function with a GP allows one to predict unseen values of the function and to calculate where it is most benefitial to evalute the function next.

We gave a GP prior for the objective function 
\begin{equation}\label{eq:objective_gp_prior}
 g(\bm\theta) \sim \mathcal{GP}(0,K(\bm\theta,\bm\theta^\prime)),
\end{equation}
where we constructed the covariance function as a sum of three components, $K(\bm\theta,\bm\theta^\prime) = K_{\text{isotr}}(\bm\theta,\bm\theta^\prime)+K_{\text{ARD}}(\bm\theta,\bm\theta^\prime)+K_{\text{sq}}(\bm\theta,\bm\theta^\prime)$ including %
an isotropic Matérn $\nu=5/2$ covariance function $K_{\text{isotr}}$, %
a Matérn $\nu=5/2$ covariance function with separate length scale for each parameter (an automatic relevance detection (ARD) covariance),%
and covariance function $K_{\text{sq}}$ derived from a quadratic polynomial mean function $\sum_{m=1}^d [1,\theta_m,\theta_m^2]\beta$, where each coefficient $\beta_m$ had an independent normal prior $\beta_m \sim N(0,\sigma_i^2)$ \citep{Rasmussen+Williams:2006}.
We gave the magnitude hyperparameter (the variance) a (half-) gaussian prior $N_+(0,1)$ for the isotropic covariance function and half Student-$t$ prior with four degrees of freedom, $t_+{\nu=4}$, for the other covariance functions. 
For the length scale of the isotropic covariance function we gave half Student-$t$ with four degrees of freedom , while in the ARD covariance function each length scale had a rior $1/l_i \sim t_{\nu=4}$.

We initialized the BO algorithm with a space filling design for the parameter values. 
We constructed the design by using $n_{\text{init}}=50$ points from a Sobol sequence \citep{sobolUniformlyDistributedSequences1976} then scaling and shifting the points so that they covered a hypercube in the parameter space within four prior standard deviations from the prior mean.
We then evaluated $g(\theta)$ at these design locations and collected these evaluations of our objective function in $\bm\Theta_0=\lbrace\bm\theta^{(0)}_1,\dots,\bm\theta^{(0)}_{n_{\text{init}}}\rbrace$ and $\mathbf{g}_0=[g(\bm\theta^{(0)}_1),\dots,g(\bm\theta^{(0)}_{n_{\text{init}}})]$. 
The initial limits for the BO were set slightly larger than the span of the initial design.

As is standard in many BO algorithms, we used the GP model to predict the posterior distribution for the objective function and to calculate the Expected Improvement (EI) at a potential new query location $\bm\theta$.
The expected improvement is defined as
\begin{equation}
\text{EI}(\bm\theta|\bm\Theta_q,\mathbf{g}_q) = \text{E}[\text{max}(\hat g-g(\bm\theta),0)], 
\end{equation}
where $\hat g$ is the current minimum of the objective function and the expectation is taken over the GP posterior predictive distribution for the objective function
\begin{equation}
g(\bm\theta)|\bm\Theta_q,\mathbf{g}_q \sim N\left(\mu(\bm\theta|\bm\Theta_q,\mathbf{g}_q), \sigma^2(\bm\theta|\bm\Theta_q,\mathbf{g}_q)\right),
\end{equation}
where $\mu(\bm\theta|\bm\Theta_q,\mathbf{g}_q)$ and $\sigma^2(\bm\theta|\bm\Theta_q,\mathbf{g}_q)$ are the posterior predictive mean and variance functions \citep{Rasmussen+Williams:2006} given the queries acquired so far, $\{\bm\Theta_q,\mathbf{g}_q\}$.

The EI can be expressed in closed form as \citep{jones1998efficient}
\begin{equation}\label{eq:EI}
\text{EI}(\bm\theta|\bm\Theta_q,\mathbf{g}_q) = (\hat g-\mu)\Phi\left(\frac{\hat g-\mu}{\sigma}\right)+\sigma\phi\left(\frac{\hat g-\mu}{\sigma}\right)
\end{equation}
where $\Phi$ is the standard normal cumulative distribution function and $\phi$ is the standard normal density function.

Since the gradient of EI is also available in closed form, we used an interior point gradient descent algorithm \citep{byrd1999interior} to find parameterizations that (locally) maximize EI. 
To find find global maxima, we repeated the gradient descent for
$n^{\ast}=10^4$ initializations from a Sobol sequence.  

In the basic implementation of BO, the objective function is next evaluated at the parameter value $\hat{\mathbf{\theta}}$ that maximizes EI.
However, as we used a computer with 20 CPU cores (and 197 GB of memory) to do the calculations, we evaluated the objective function at $n_{\text{batch}}=20$ local maxima in parallel at each iteration of the BO algorithm (see Section~\ref{sec:batch_optimization} for details).
After calculating the objective function at these $n_{\text{batch}}$ query points we appended the training data with the respective values so that $\bm\Theta_{q+1}=\bm\Theta_{q}\cup \lbrace\bm\theta^{(q+1)}_1,\dots,\bm\theta^{(q+1)}_{n_{\text{batch}}} \rbrace$ and $\mathbf{g}_{q+1}=[\mathbf{g}_{q}^T, g(\bm\theta^{(q+1)}_1),\dots,g(\bm\theta^{(q+1)}_{n_{\text{batch}}})]^T$. 

We repeated the above steps of fitting the emulator to training data from the objective function~\eqref{eq:objective}, locating the local maxima of EI-function, and evaluating new training data with objective function, until convergence criteria were met, or we reached the end of our iteration budget, $N_\text{itermax}$. 
Let $(\hat{\bm\theta}, \hat g)$ and $(\tilde{\bm\theta}, \tilde g)$ be the two parametrisations with the smallest function evaluations seen so far.
We considered the optimization converged, when we had evaluated at least $N_{\text{itermin}} = 5$ batches and any of the following criteria were met:
\begin{itemize}
\item $\Vert \tilde{\bm\theta} - \hat{\bm\theta} \Vert \leq \epsilon_{\theta}$ and $\vert\tilde g - \hat g\vert \leq \epsilon_g$
\item $\text{EI}(\bm\theta_b|\bm\Theta_q,\mathbf{g}_q) \leq \epsilon_\text{EI}, \forall b=1,\dots,n_\text{batch}$
\item $q \geq N_\text{itermax}$
\end{itemize}
The convergence limits were set to $\epsilon_\theta=10^{-2}$, $\epsilon_g=10^{-2}$, $\epsilon_{\text{EI}}=10^{-3}$.

Once we reached the end point of log minus log posterior optimization, we constrained the search space (see Section~\ref{sec:constrain_cand}) and repeated the above optimization steps for the negative log posterior density within the new limits.
Typically, the optimization had already converged close to the final mode when optimizing the log-minus-log posterior density, so that we only needed to evaluate a small number of batches in this second phase.

When the optimization is close to convergence, the covariance matrix of the GP emulator may become close to singular. 
This happened typically in the log minus log phase, where the much smoother surface can occasionally lead to the GP having long length scales and high correlations between points. 
If the covariance matrix was about to become singular after including a candidate point into a batch, we interrupted the batch acquisition strategy and used a smaller batch. 
The BO algorithm and the GP emulators were implemented with Matlab toolbox GPstuff \citep[][]{Vanhatalo+Riihimaki+Hartikainen+Jylanki+Tolvanen+Vehtari:2013}.

\subsubsection{Decreasing the variability of the objective function}\label{sec:decreaseVariability}

The GP emulators have trouble in simultaneously fitting to large and small variation in the objective function.
Hence, since the aim of BO algorithm was to find the mode, we truncated the objective function~\eqref{eq:objective} values that corresponded to orders of magnitude smaller posterior densities than the density at the mode before training the GP emulator.
At each iteration, given the value of the objective function at the current optimum, $\hat g$, the largest values of the objective function were truncated so that  $g>\hat g+10$ were set to $\hat g+10$.
We chose the threshold $10$ based on the log density of the Gaussian distribution, for which this threshold corresponds approximately to the highest $99.9\%$ probability mass region (see Section \ref{sec:constrain_cand}).

\subsubsection{Batch selection of query points}\label{sec:batch_optimization}

Intuitively we should generalize the EI function so that it maximizes the joint information from the multiple query points \citep{Gonzales+etal:2016,Wang+etal:2018,Zhan+Xing:2020}.
However, these algorithms are computationally intensive and complicated to tune to work well. 
Our main objective in choosing a batch design was to leverage the benefit from free CPU cores with minimal extra computation and programming.
Hence, we aimed at finding a batch of good, but not necessarily optimal, candidates that provided extra information and, thus, speeded up the optimization.

With these criteria in mind, we used a strategy based on the heuristic strategies in \citep{ginsbourgerKrigingWellSuitedParallelize2010}.
This is a (pseudo) sequential strategy, where we select $n_\mathrm{batch}$ query points one at a time by maximizing EI conditional on previous evaluations. However, instead of immediately evaluating the objective function at each query point, which would revert the batch optimization to the traditional sequential optimization, we trained the GP using fast to evaluate surrogate values $\tilde g$. We also did not update the emulator's hyperparameters while selecting query points.

An intuitive choice for the $\tilde g$ would be to use the posterior predictive mean $\mu(\tilde{\bm\theta}|\bm\Theta,\mathbf{g})$ since doing so would retain the same mean function for the GP, while reducing uncertainty near the chosen points. \citet{ginsbourgerKrigingWellSuitedParallelize2010} call this strategy Kriging Believer.
However, they found that when the GP predicts expectations below the currently seen minimum, $\hat{g}$, this strategy can lead to the batch clustering around the same query point. 
To avoid this, they tried another strategy, Constant Liar, where they use a constant $L$ for all surrogate values and found that $L = \hat{g}$ performed the best.

Our strategy was a combination of these two strategies. When the GP predictive mean was larger (i.e., worse) than the current optimum, $\hat g$, we used the predictive mean as a surrogate value $\tilde{g}$. When the GP predictive mean was smaller (i.e., better) than the current optimum, we used the current optimum as a surrogate value. As a result, for majority of the batch query points the behaviour of our batch selection was the same as that of the Constant Liar, since local maxima of the EI function tended to have smaller GP predictive means than $\hat g$. 
However, occasionally points with $\mu(\tilde{\bm\theta}|\bm\Theta,\mathbf{g}) > \hat{g}$ were selected. These points typically corresponded to local maxima where the GP predictive variance was high. Using the mean as a surrogate in these cases did not lead to clustering, but instead repelled clustering more strongly than if we used $\tilde g = \hat{g}$.
We have summarised our strategy in Algorithm~\ref{alg:batchAcq}.

\begin{algorithm}
 \caption{Algorithm for selecting batch query points. Returns the next $n_\mathrm{batch}$ query points where the objective function is to be evaluated. As inputs, gets the previous query points $\bm\Theta^{(q)}$ and their corresponding objective function values $\mathbf{g}^{(q)}$ from previous $q$ batches.}\label{alg:batchAcq}
 \begin{algorithmic}
   \Function {batchAcquisition}{$\bm\Theta^{(q)}$, $\mathbf{g}^{(q)}$, $n_\mathrm{batch}$}
    \State $\hat{g} \gets \min\mathbf{g}^{(q)}$
    \State $\tilde{\bm\Theta} \gets \bm\Theta^{(q)}$
    \State $\tilde{\mathbf{g}} \gets \mathbf{g}^{(q)}$
    \For {$i \gets 1 \dots n_\mathrm{batch}$}
      \State $\tilde{\bm\theta}^{(q+1)}_i = \argmax_{\bm\theta}\mathrm{EI}(\bm\theta | \tilde{\bm\Theta}, \tilde{\mathbf{g}})$
      \State $\tilde{g}^{(q+1)}_i = \max\left\lbrace \mu(\tilde{\bm\theta}^{(q+1)}_i|\tilde{\bm\Theta}, \tilde{\mathbf{g}}), \hat{g} \right\rbrace$
      \State $\tilde{\bm\Theta} \gets \tilde{\bm\Theta} \cup \left\lbrace \tilde{\bm\theta}^{(q+1)}_i \right\rbrace$
      \State $\tilde{\mathbf{g}} \gets \tilde{\mathbf{g}} \cup \lbrace\tilde{g}^{(q+1)}_i\rbrace$
    \EndFor
    \State \textbf{return} $\lbrace \bm\theta^{(q+1)}_i, \dots, \bm\theta^{(q+1)}_{n_\mathrm{batch}}  \rbrace$
   \EndFunction
 \end{algorithmic}
\end{algorithm}

\subsubsection{Constraining search space and optimizing log-posterior}\label{sec:constrain_cand}

After finding the minimum of the log-minus-log posterior, we constrained the candidate space to a subset of the original search space consisting of the smallest hypercube containing all non-implausible points similarly to \citet{williamson2013history}.
We defined a parametrisation to be non-implausible if it was within the $90\%$ highest posterior density (HPD) region with at least $5\%$ probability (i.e. points outside the HPD region with probability over $95\%$ are implausible).
The HPD region was defined by approximating the true posterior density near the mode with a multivariate Gaussian with mean $\bm\mu$ and covariance $\bm\Sigma$. For a $d$ dimensional multivariate Gaussian the HPD region is \citep{ruben1960probability}
\begin{equation}
\Omega_\text{HPD}(p) = \{\bm\theta : (\bm\theta-\bm\mu)^T\bm\Sigma^{-1}(\bm\theta-\bm\mu) \leq (\chi_d^2)^{-1}(p)\},
\end{equation}
where $(\chi_d^2)^{-1}(p)$ is the inverse cumulative distribution function of the $\chi^2$ distribution with $d$ degrees of freedom. 
Hence, assuming that the posterior density $p(\bm\theta|\mathbf{y})\propto p(\bm\theta|D)p(\bm\mu|D)$ is close to Gaussian, this suggests a heuristic criteria based on difference in log densities from the posterior mode, $\hat{\bm\theta}$, such that
\begin{align}
\hat\Omega_\text{HPD}(p) &= \left\lbrace \bm\theta : 2 \left( \ln p(\bm\theta|\mathbf{y}) - \ln p(\hat{\bm\theta}|\mathbf{y}) \right) \leq (\chi_d^2)^{-1}(p) \right\rbrace \nonumber \\ 
&= \left\lbrace \bm\theta : \ln p(\bm\theta) + \mathcal{L}(\bm\theta|\mathbf{y}) \leq \ln p(\hat{\bm\theta}) + \mathcal{L}(\hat{\bm\theta}|\mathbf{y}) +\frac{1}{2}(\chi_d^2)^{-1}(p) \right\rbrace.
\end{align}
After BO had converged, we had an estimate for the posterior mode and the unnormalized log posterior density at it, $\ln p(\hat{\bm\theta}) + \mathcal{L}(\hat{\bm\theta}|\mathbf{y})$. 
We could then use the GP emulator to approximate the probability that any $\bm\theta$ belongs to the above defined HPD region
\begin{equation}\label{eq:pr_hpd_approx}
\Pr(\bm\theta \in \Omega_\text{HPD}(p)) \approx \Phi \left( \frac{\ln p(\hat{\bm\theta}) + \mathcal{L}(\hat{\bm\theta}|\mathbf{y})+\frac{1}{2}(\chi_d^2)^{-1}(p)-\mu(\bm\theta|\bm\Theta,\mathbf{g})}{\sigma(\bm\theta|\bm\Theta,\mathbf{g})} \right).
\end{equation}

We then estimated the non-implausible hypercube by generating $N_{\text{constr}} = 10^5$ quasi random points from the Sobol set \citep{sobolUniformlyDistributedSequences1976, joeRemarkAlgorithm6592003} within the current candidate boundaries, computed the GP emulated log minus log posterior
\eqref{eq:objective_gp_prior} at them, and constrained our candidate space (i.e., the range of FICOS parameter values from where the posterior mode was searched) to the smallest hyper cube containing all points for which the probability of being within the $90\%$ HPD was above $5\%$.

We also used wider buffer region $\Omega_\text{buffer}$ corresponding to our approximation of the $99.5\%$ HPD. We kept all of the current objective function evaluations $\mathbf{g}$ within this region for training our GP emulator, while looking for new candidates in the more constrained $90\%$ HPD.
The motivation for this buffer is to achieve a better GP fit within the HPD region (see also Section~\ref{sec:decreaseVariability}). 
If we trained our GP with evaluations far away from the mode with posterior densities close to $0$, these points would have an effect on the GP hyperparameters and lead to worse fit near the mode.
Conversely, if we trained our GP with only the evaluations within the $90\%$ HPD, we might discard evaluations that informed the newly constrained boundary and would need to evaluate additional candidates near the boundary due to the increased uncertainty.

\subsection{Posterior distribution for calibration parameters}\label{sec:GPpostEmul}
  
After the BO for the log posterior converged, we estimated the log (unnormalized) posterior density for the calibration parameters using the GP emulator as in the BO-algorithm.
To ensure good coverage of training points for the GP emulator within the HPD region (see Section~\ref{sec:constrain_cand}) we constructed a space filling design over it.
To do this, we first constructed a multivariate Gaussian approximation for the posterior density around the mode and used this approximation to quide the construction of the space filling design \citep[see, e.g.,][for a similar apporoaches]{kuoQuasiMonteCarloHighly2008,Vanhatalo+etal:2010,dickHighdimensionalIntegrationQuasiMonte2013}. 

First, we calculated the Laplacian of the GP mean function at the posterior mode, $\hat{\theta}$, and inverted it to get an estimate of the covariance matrix
\begin{equation}
 \hat{\bm{\Sigma}} = \left(\Delta \mu(\hat{\bm{\theta}}|\Theta_q,\mathbf{g}_q) + \delta \mathbf{I}\right)^{-1} = \mathbf{V} \bm\Lambda \mathbf{V}^T
\end{equation}
where $\bm\Theta_q,\mathbf{g}_q$ are all the query data from the BO for the log posterior and $\delta \mathbf{I}$ is an optional jitter term. This term can be added to the diagonal of the Laplacian if it is close to singular, although we did not need to do so for the results presented in here.
We then calculated the eigen decomposition $\mathbf{V} \bm\Lambda \mathbf{V}^T$ of $\hat{\bm{\Sigma}}$ and constructed a space filling design $\bm{Z}$ along the eigen vectors $\bm{V}$
\begin{equation}
 \bm{Z} = \hat{\bm{\theta}} + r_\text{spfill}\bm{W}\bm{V}\sqrt{\bm\Lambda}
\end{equation}
where $\bm{W}$ is a space filling design of $n_{\text{spfill}}$ Sobol sequence points within $[-1,1]^d$ and $r_\text{spfill}$ is a length parameter controlling the volume of the design.

We adjusted the volume and density of the design as a compromise between computational budget and coverage. We considered the smoothness of the posterior density within the HPD based on the emulator length scales, and found that we could cover approximately $99\%$ of the probability mass region while interpolating accurately with length $r_\text{spfill} = \frac{1}{2}(\chi_5^2)^{-1}(0.99)\approx 7.5$ and $n_{\text{spfill}}=500$ design points.

After evaluating the log posterior at the space filling design points and appending the evaluations to the earlier data,
i.e. $\bm\Theta_{\text{final}}=\bm\Theta_q\cup \bm\Theta_{\text{spfill}},\mathbf{g}_{\text{final}}=\mathbf{g}_q \cup \mathbf{g}_{\text{spfill}}$,
we trained the GP emulator for one final time.
We then drew samples from the (approximate) posterior distribution with slice sampling \citep{Neal:2003} using the GP predictive mean $\mu(\bm\theta|\bm\Theta,\bm g)$ as the target (log) density.

\subsection{Posterior predictive inference and scenario predictions}

One of the main uses of the FICOS model is to make scenario predictions with alternative nutrient loads to help management and policy planning.
Hence, we wanted to propagate the calibration parameter uncertainty into the FICOS model predictions. 
We did this by running the FICOS model with the posterior samples of the calibration parameters.
We selected $n_{\mathrm{MCMC}} = 200$ of these samples with large enough thinning so that the autocorrelation of the samples was practically zero.
We then run all of the FICOS scenarios with these 200 (approximately) independent samples, $\bm\theta_{s}$, to obtain Monte Carlo estimate for the posterior distribution for the scenario predictions.

Scenario predictions are typically compared to a business as usual (BAU) scenario. Posterior samples for BAU correspond to FICOS runs in parameter calibration phase, so we had access to $f(\theta_{s})$, which allowed us to also sample from the posterior distribution of the noice variances and through that from the predictive distribution for (new) observations $\tilde{\mathbf{y}}_{j,k}$. The posterior predictive distribution for $\tilde{\mathbf{y}}_{j,k}$ is given by
\begin{equation}\label{eq:postPredObs}
p(\tilde{\mathbf{y}}_{j,k} | \mathbf{y}) = \int_{\bm\theta}\int_{\sigma^2_{j,k}} p(\tilde{\mathbf{y}}_{j,k}|\bm\theta, \sigma^2_{j,k}) p(\sigma^2_{j,k}|\bm\theta,\mathbf{y}_{j,k}) p(\bm\theta |\mathbf{y}) d\sigma^2_{j,k} d\bm\theta.
\end{equation}
We drew samples from \eqref{eq:postPredObs} by drawing samples from each of its constituent components. For the innermost component, $\bm\theta|\mathbf{y}$, we use thinned slice samples $\bm\theta_s$ drawn earlier. Then, for each $\bm\theta_s$, we sampled noise variances $(\sigma^2_{j,k})_s | \bm\theta_s, \mathbf{y}$ from the scaled Inv-$\chi^2$ distribution, which we explain in more detail below. 
Finally, we sampled $(\tilde{\mathbf{y}}_{j,k})_s | \bm\theta_s, (\sigma^2_{j,k})_s \sim N(\mathbf{f}_{j,k}(\bm\theta_s) | (\sigma^2_{j,k})_s)$ for each $s = 1,\dots,n_\mathrm{MCMC}$.

We omit the indices $j,k$ for the rest of this section for more convenient notation.
When none of the observations were censored, we could sample directly from
\begin{equation}\label{eq:conditionalPosteriorSigma}
\sigma^2 | \mathbf{y}, \bm\theta_s \sim \mathrm{scaled\,Inv}-\chi^2 
\left(\nu_0+n \,, \frac{\tau_0^2+ \bm\epsilon_s^T \bm\epsilon_s}{\nu_0+n}\right),
\end{equation}
where $\bm\epsilon_s = \mathbf{y}-\mathbf{f}(\bm\theta_s)$ and $n$ is the number of observations. 

When some of the observations were censored, we used Gibbs sampling \citep[][]{gelfandIllustrationBayesianInference1990},
where we alternated between sampling the unobserved values for censored measurements $\mathbf{y}_c \leq \mathbf{a}$ and noise variances $\sigma^2$ conditional on the other parameters.
That is, we initialized
$\sigma^2 = \tau_0^2$, then sampled each $(y_c)_{j}$, $j=1,\dots,n_{\text{censored}}$ from the truncated normal with inverse CDF sampling:
\begin{align*}
u_{j} &\sim \mathrm{Unif}(0,1) \\
(y_c)_{j}| \sigma^2, \bm\theta_s &= f_j(\bm\theta_s) + \sigma^2 
\Phi^{-1}\left( u_{j}
\Phi ( \frac{a_j - f_j(\bm\theta_s)}{\sigma^2} ) \right),
\end{align*}
and then, with $\epsilon = \begin{bmatrix}
(\mathbf{y}_c)\\ \mathbf{y}_p
\end{bmatrix}- \mathbf{f}(\bm\theta_s)$, sampled the noise variances:
\begin{align*}
\sigma^2 | (\mathbf{y}_c), \mathbf{y}_p, \bm\theta_s &\sim \mathrm{scaled\, Inv}-\chi^2 \left( \nu_0+n, \frac{\tau_0^2+
\epsilon^T 
\epsilon}{\nu_0 + n} \right).
\end{align*}
These steps were repeated until convergence and appended the last sample of $\sigma^2$ with $\bm \theta_s$.

The FICOS runs with the BAU scenario predictions also allowed us to calculate the true posterior density, $p(\bm\theta_s|\mathbf{y})$, for each posterior sample, $\bm\theta_s$, used in the scenario predictions. 
We used these true posterior densities to calculate importance weights for the posterior samples, which we then used to attain an importance weighted Monte Carlo estimate for the scenario predictions. 
That is, we calculated importance weights $w_s = p(\bm\theta_s|D) / \exp (-\mu(\bm\theta_s|\bm\Theta,\bm g))$ for the predictive samples to correct for possible bias in the sampling using the GP emulator.

\section{Results and Discussion}

\subsection{Batch Bayes optimization}

\begin{figure}[t]
\centerline{
 \includegraphics[width=3in]{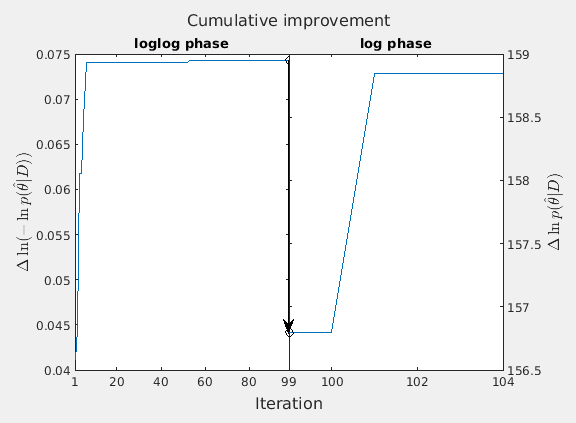} 
 }
 \caption{Cumulative improvement over the initial mode $\hat{\bm\theta}_0$ of the objective function at each iteration of the BO algorithm.}
 \label{fig:cumulativeImp}
\end{figure}

We tested our approach for approximate Bayesian inference on FICOS model parameters by running the inference multiple times, first with three calibration parameters and later with the five calibration parameters outlined above.
We also varied the size of the space filling initialization. 
In all these tests, the batch BO algorithm always converged reliably. 
We also verified the performance of our batch BO strategy by plotting the cumulative improvement over the initial mode $\hat{\bm\theta}_0$ (see Figure~\ref{fig:cumulativeImp}). For the query points we plotted all 10 pairwise scatter plots at each iteration to verify visually that the query points continue to explore the parameter space after the initial space filling design. We included a small subset of these plots into Figure~\ref{fig:batchScatter} (see Appendix).

Based on these results, the batch BO algorithm efficiently explored the parameter space in the early stage of the algorithm and located the modal region fast.
After this the algorithm spent quite some number of iterations fine tuning the modal estimate before the convergence criteria for the first stage, log minus log posterior, objective function were met. 
In the second stage, negative log posterior, the optimization converged in minimum number of iterations.
The amount of optimization budget spent in the first stage optimization and the second stage optimization depends on the convergence criteria for the BO algorithm. 
We found that using relatively strict convergence criterion in the first stage (Figure~\ref{fig:cumulativeImp}) worked better than trying to locate the first stage log minus log posterior mode roughly and spending more time on optimizing the log posterior. 
Because of the truncation in our objective function, its surface is mostly flat with sharp peaks near the mode. 
\citet{dewanckerStratifiedAnalysisBayesian2016} calls these types of functions "mostly boring" and found that GP based BO algorithms perform the best in optimizing them.
The sharp peaks may still cause problems for the GP emulator during the log posterior phase, however, especially if we have only a small amount of function evaluations around the mode due to most of the initial design points being outside of the HPD region.
Hence, we found it more efficient to spend more of the computational budget in the log minus log phase with much smoother surface, where we could refine the GP emulator and constrain the search space, before moving on to the second phase.

Good computational performance for the posterior inference for the calibration parameters, including the BO (Section~\ref{sec:BO}) and posterior sampling (Section~\ref{sec:GPpostEmul}) required typically approximately 2000 FICOS model runs in total. 
Out of these, 50 model runs were needed for the initial design of BO and 500 for the final space filling design.
The rest of the FICOS model runs, approximately 1500 model runs, were used in batch BO algorithm corresponding to roughly 100 iterations for the convergence.
While most of these batches had the maximum batch size of 20, some batches were smaller due to numerical issues in the batch candidate selection algorithm (see Section~\ref{sec:batch_optimization}).
Since each FICOS model run took in average 4 minutes, the full inference was done in approximately 7 hours.
The results for the BO convergence for the specific initialization detailed in Section~\ref{sec:BOsummary} is shown in Figure~\ref{fig:cumulativeImp}.

The algorithm is fairly simple, however, which is why it is plausible that the search for the posterior mode could be made more efficient with more advanced selection of batch members, for example, via local penalization approach \citep{Gonzales+etal:2016}.
An alternative to batch BO would be to select the query points using non-myopic strategies which aim to maximize the joint information from multiple sequential query points \citep{Xubo+etal:2020,Jiang+etal:2020}.
However, in our application the batch strategy seems more important than correcting for the nonmyopicity since the bottle neck is the time needed for one objective function evaluation whereas we can efficiently parallelize these evaluations.

Moreover, the most important future development for our approach would be to scale it up to work efficiently for higher resolution FICOS models. 
In this work, we used the smallest possible spatial resolution for FICOS for computational reasons. 
However, it would be desirable to calibrate the FICOS model in its highest spatial resolution, corresponding to nautical mile, since that is the resolution at which many of the management and policy analyses are done. 
Since running the FICOS model with its highest resolution takes from one to few days, our approach does not scale well for it yet. 
We believe the most promising approach to increase the resolution of the calibrated FICOS model would be to use multifidelity BO \citep[e.g.,][]{Wu+etal:2020,Zanjani+etal:2023} to locate the posterior mode of the calibration parameters.
With multifidelity BO approach, we could attain improved calibration by combining high and low resolution FICOS runs directed by efficient use of the total computational budget.

\subsection{Posterior inference}

\begin{figure}
 \includegraphics[width=4.3in]{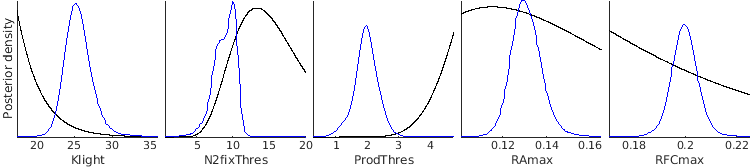}
 \caption{Parameter posterior marginal densities. The blue line is the posterior density and the black line is the prior density scaled to match the posterior density.%
 }
 \label{fig:marginalPosterior}
\end{figure}

The posterior distributions for the five calibration parameters of the BGC model differed clearly from the respective prior distributions while being still biologically highly reasonable (Figure~\ref{fig:marginalPosterior})
This indicates that the monitoring data was highly informative on the calibration parameters. The light-related parameters showed small pairwise correlations as can be expected, but otherwise these parameters were not correlated (Figure~\ref{fig:jointPosterior}).

\begin{sidewaysfigure}[p]
	\includegraphics[width=\textheight,trim={0 0 0 20px},clip]{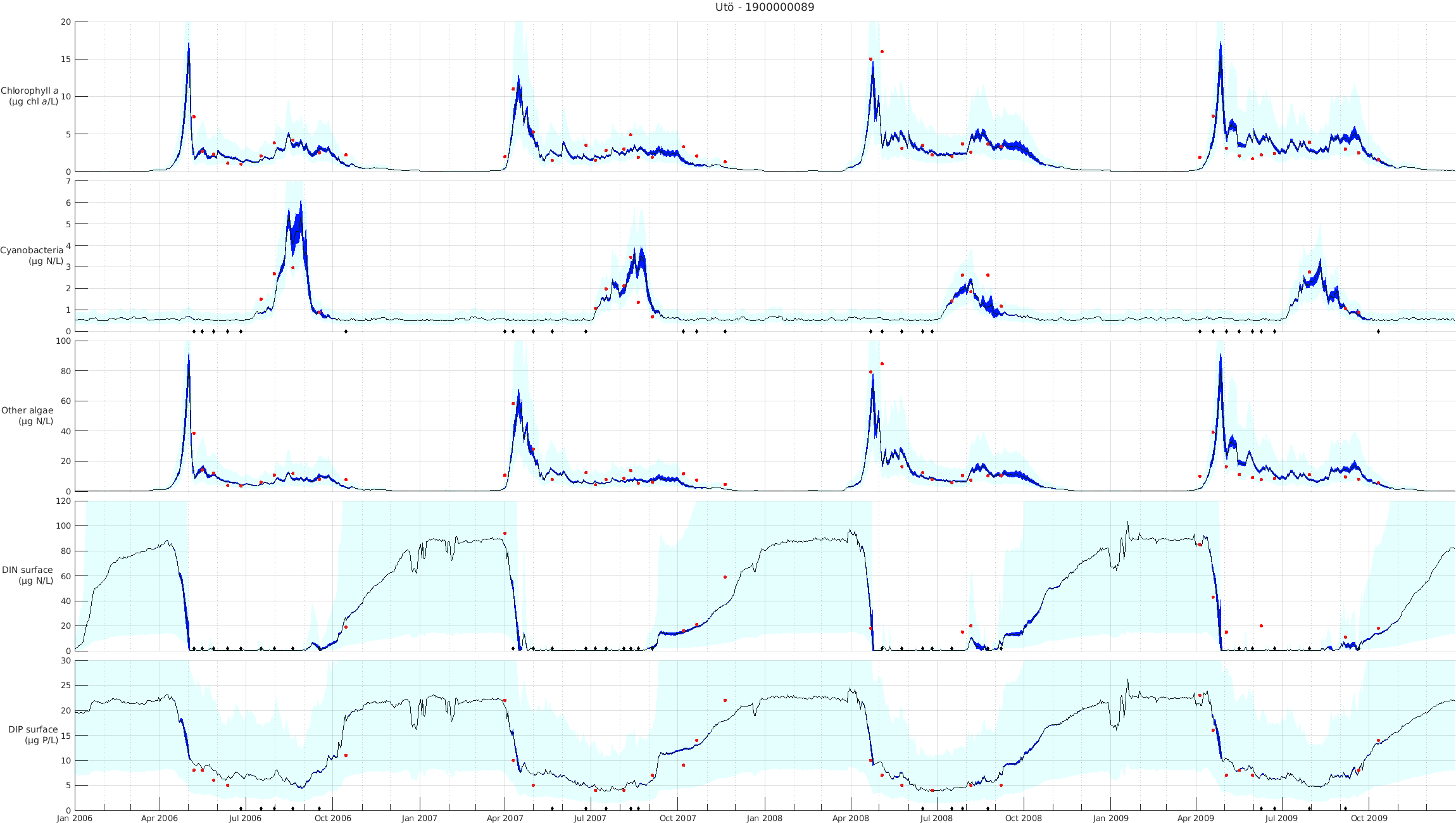}		
\caption{Posterior predictive time series for the intensive station in Utö (black line) with the associated 95\% credible interval propagated from the posterior of the calibration parameters (dark blue intervals) for the calibration time period 2006-2009. The light blue intervals show the 80\% credible interval for predicted observations. The red dots show the non-censored calibration data from the same station and black diamonds mark the censored observations (Section~\ref{sec:lik_censored}).}
\label{fig:time_series}
\end{sidewaysfigure}

The posterior distributions for the time series in the intensive monitoring stations showed clear seasonal patterns and matched well the water quality measurements of the training time interval from 2006 to 2015 (see Figure~\ref{fig:time_series} and Table~\ref{table:R2_median}).
The explanatory power of the posterior mean of the FICOS model output, as measured by the $R^2$ statistics, was generally over 0.8 for all the calibration variables (DIN, DIP and algae densities), which can be considered as a good performance for a hydrological model of this complexity \citep{n.moriasiHydrologicWaterQuality2015}. 

The posterior uncertainty in FICOS model output was, however, rather small (Figure~\ref{fig:time_series}).
This is reasonable since the spatial resolution of the FICOS model was in the sub-basin scale at which the FICOS model outputs are much smoother than they would be with higher spatial resolution. 
Moreover, our observation model~\eqref{eq:obs_model} did not account for possible temporal correlations in the measurements, leading to a large number of effective observations in the likelihood function relative to the number of calibrated parameters.
Including these potential correlations in the observation model would make it more realistic and is, thus, a natural future development for calibrating the FICOS model.

\begin{table}[h]
\begin{center}
\begin{tabular}{|c| c c c c c c c|}
\hline
Station/Variable & DIN1 & DIP1 & DIN2 & DIP2 & chla & A & FC \\ [0.5ex]
\hline
Utö & 0.55 & 0.98 & 0.99 & 0.99 & 0.76 & 0.93 & 0.68 \\
Seili &	0.34 & 0.88 & 0.82 & 0.95 & 0.81 & 0.94 & 0.61 \\
Brändö & 0.91 & 0.99 & 0.98 & 0.97 & 0.52 & 0.90 & 0.85 \\ [1ex]
\hline
\end{tabular}
\caption{Coefficient of determination (R$^2$) for each posterior mean timeseries prediction for (non-censored) observations of log nutrient, chlorophyl a, and algae concentrations. Since the simulator was known not to accurately model winter and early spring nutrients in Seili with the resolution used in this study, the nutrient R$^2$ for Seili was calculated only for the summer months (from June to September).}
\label{table:R2_median}
\end{center}
\end{table}

\subsection{Scenario predictions}\label{sec:posterior_predictions}

To demonstrate the use of the FICOS model in management and policy planning applications, and to highlight the benefits from our approximate Bayesian inference algorithm outlined above, we conducted four scenario predictions for the chlorophyll $a$ concentration in the Archipelago Sea: 
Business as usual, corresponding to a scenario where catchment area nutrient load does not change from its historical level as well as 20\%, 40\%, 60\% and 80\%  reduction in catchment area nutrient load. 
Each of the scenario predictions was conducted by rerunning the FICOS model for years 2006-2015 so that all other inputs were the same as in the parameter inference phase but the catchment area nutrient load was altered.
We then calculated the probability to achieve the Good Ecological Status (GES) in the inner Archipelago Sea as defined in the European Union (EU) Water Framework Directive (WFD) for each of the scenarios.
In the inner Archipelago Sea, the GES was attained when the average chlorophyll $a$ concentration between 1 July and 7 September is below 3.0 $\mu$g/l \citep{Aroviita+etal:2019}.

The scenario predictions for the average chlorophyll~$a$ concentration differed clearly from each others and according to them the good ecological status in the inner Archipelago Sea is not attainable with catchment area nutrient load reductions only. All the scenario predictions contained large amounts of uncertainty, with business as usual scenario having the largest uncertainty and uncertainty decreasing as the nutrient loads decrease.

\begin{figure}[t]
\vspace{.3in}
\centerline{
	\includegraphics[width=4in]{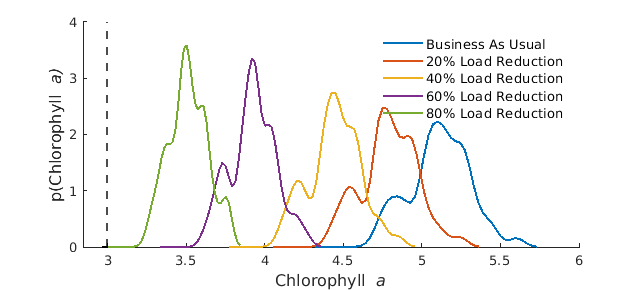}
	}
\caption{Mean Cholorophyll-$\textit{a}$ posterior predictive densities in the inner archipelago under nutrient load reduction scenarios with decreasing catchment area nutrient loadings.}\label{fig:scenario_predictions}
\end{figure}

\clearpage
\section{Conclusions}

Complex computer models are used in many fields of science, industry, and public governance. 
The general challenge with using them is how to quantify the uncertainties related to their results. 
Main sources of uncertainty in deterministic simulators include the parameter uncertainty, model inadequacy, residual variation, and observation error in the data \citep{Kennedy+OHagan:2001,Kennedy+etal:2007,Volodina+Challenor:2021}. 
In this work, we focused on estimation and uncertainty assessment of five key unknown biological parameters of the FICOS nutrient load model (\ficosManuscript) as well as on the residual variation, and observation error in the calibration data.
We did not, however, account for the model inadequacy and our treatment of the residual variation and observation error were rather simplistic. 
For example, we did not include spatial nor temporal correlations to our residual model.

Despite these evident future development needs, our approach was able to produce significant improvement to the current state of the nutrient load modeling. 
The computational methods described in this work were developed in parallel with the development of the FICOS BGC model compartment. 
The inference algorithm presented here had important role also in guiding the development of the FICOS model formulations by allowing statistically sound testing of alternative model formulations through formal model fitting and assessment.
Moreover, the most common practice in parameterizing these models has been to use theoretical considerations, expert assessments, and visual inspection of the model fit to calibration data - and to ignore the uncertainty in model outputs altogether. 
By providing statistically reasoned and coherent parameter estimates, as well as uncertainties associated with them, we also promote science based decision making by increasing the credibility and transparency of this large scale nutrient impact model, that is already in operational use in water management and policy planning.

\section*{Funding sources}

This work was supported by the Research Council Finland (project no. 317255) and Jane and Aatos Erkko Foundation.

\bibliographystyle{agsm}
\bibliography{manuscript}
\newpage


\appendix

\setcounter{figure}{0}

\section{Log likelihood for censored observations}\label{sec:appendix_censored_loglik}

First, we denote the central $n$-dimensional multivariate normal (MVN) integral by 
\begin{equation}\label{eq:mvn_integral}
\bm\Phi_n(\mathbf{b}|\bm\Sigma) = \frac{1}{\sqrt{|\bm\Sigma|(2\pi)^n}}\int_{-\infty}^{b_1} \dots \int_{-\infty}^{b_n} \exp\left(-\frac{1}{2} \mathbf{x}^T \bm\Sigma^{-1} \mathbf{x}\right) d\mathbf{x}.
\end{equation}

Now, let $\textbf{y} \sim \mathbf{t}_n(\bm\mu, \bm\Sigma, \nu)$, where $\mathbf{t}_n$ is the $n$ dimensional multivariate $t$ (MVT) distribution with location vector $\bm\mu$, scale matrix $\bm\Sigma$, and $\nu$ degrees of freedom. Let $\mathbf{a} \in \mathbb{R}^n$ be a vector of upper bounds. The truncated likelihood \eqref{eq:lik_censored} is then a non-central MVT integral
\begin{equation} \label{eq:lik_censored_mvt}
\begin{gathered} 
\Pr( \mathbf{y} \leq \mathbf{a},|\bm\mu,\bm\Sigma) = \Pr(y_{i} \leq a_i, \dots, y_{n} \leq a_n,|\bm\mu,\bm\Sigma)\\ 
 = \int_{-\infty}^{a_1} \dots \int_{-\infty}^{a_n} \mathbf{t}_n\left( \mathbf{y} | \bm\mu, \bm\Sigma, \nu \right) d\mathbf{y} \\
 = \frac{\Gamma(\frac{\nu+n}{2})}{\Gamma(\frac{\nu}{2}) \sqrt{|\bm\Sigma|(\nu\pi)^n}} \int_{-\infty}^{a_1} \dots \int_{-\infty}^{a_n} \left(1 + \frac{(\mathbf{y}-\bm\mu)^T \bm\Sigma^{-1} (\mathbf{y}-\bm\mu ) }{\nu} \right)^{-\frac{\nu+n}{2}} d\mathbf{y}.
 \end{gathered}
\end{equation}
This form of the integral is difficult to compute directly, hence we instead transform it to a more tractable form. 

We do this by following \citep{genzComputationMultivariateNormal2009} and writing \eqref{eq:lik_censored_mvt} as an inner MVN integral and an outer line integral over the $\chi$ distribution.
First, we shift the limits by substituting $\mathbf{x} = \mathbf{y}-\bm\mu$, transforming \eqref{eq:lik_censored_mvt} to a \textit{central} MVT integral
\begin{equation}\label{eq:lik_censored_shift_limits}
\frac{\Gamma(\frac{\nu+n}{2})}{\Gamma(\frac{\nu}{2}) \sqrt{|\bm\Sigma|(\nu\pi)^n}} \int_{-\infty}^{a_1-\mu_1} \dots \int_{-\infty}^{a_n-\mu_n} \left(1 + \frac{\mathbf{x}^T \bm\Sigma^{-1} \mathbf{x} }{\nu} \right)^{-\frac{\nu+n}{2}} d\mathbf{x},
\end{equation}
which then allows us to write it using the central MVN integral \eqref{eq:mvn_integral}:
\begin{equation}\label{eq:lik_censored_mvn}
 \frac{2^{1-\frac{\nu}{2}}}{\Gamma(\frac{\nu}{2})} \int_0^\infty  s^{\nu-1}\exp \left(-\frac{s^2}{2} \right) \bm\Phi_n \left( \frac{s }{\sqrt{\nu}} \left(\mathbf{a}-\bm\mu \right)|\bm\Sigma \right) ds
\end{equation}
Finally, \eqref{eq:lik_censored_mvn} can be inverse CDF transformed into a line integral
\begin{equation}\label{eq:lik_censored_line}
\int_0^1  \bm\Phi_n \left(\frac{\chi_\nu^{-1}(t)}{\sqrt{\nu}} \left(\mathbf{a} -\bm\mu \right)|\bm\Sigma \right) dt,
\end{equation}
where $\chi_\nu^{-1}$ is the inverse CDF of the $\chi$ distribution with $\nu$ degrees of freedom.

In our case the observations are uncorrelated, since $\bm\Sigma = \tau^2 \textbf{I}_n$, which allows us to factorize the inner integral into a product of univariate integrals:
\begin{equation}\label{eq:lik_censored_mvn_univariate}
\int_0^1 \prod_{i=1}^n \Phi \left( \frac{\chi_\nu^{-1}(t)}{\tau\sqrt{\nu}} \left(a_i-\mu_i \right) \right) dt.
\end{equation}
It is possible to derive \eqref{eq:lik_censored_mvn_univariate} more directly from the definition of the (uncorrelated) MVT distribution. However, the more general form in \eqref{eq:lik_censored_line} allows us to expand to more complex observation models, such as those that account for temporal correlation.


\section{Extra results}
\setcounter{figure}{0}

\begin{figure}[h!b]
 \begin{subfigure}[b]{0.495\textwidth}
  \centering 
  \includegraphics[trim={0 5px 0 20px},clip,width=\textwidth]{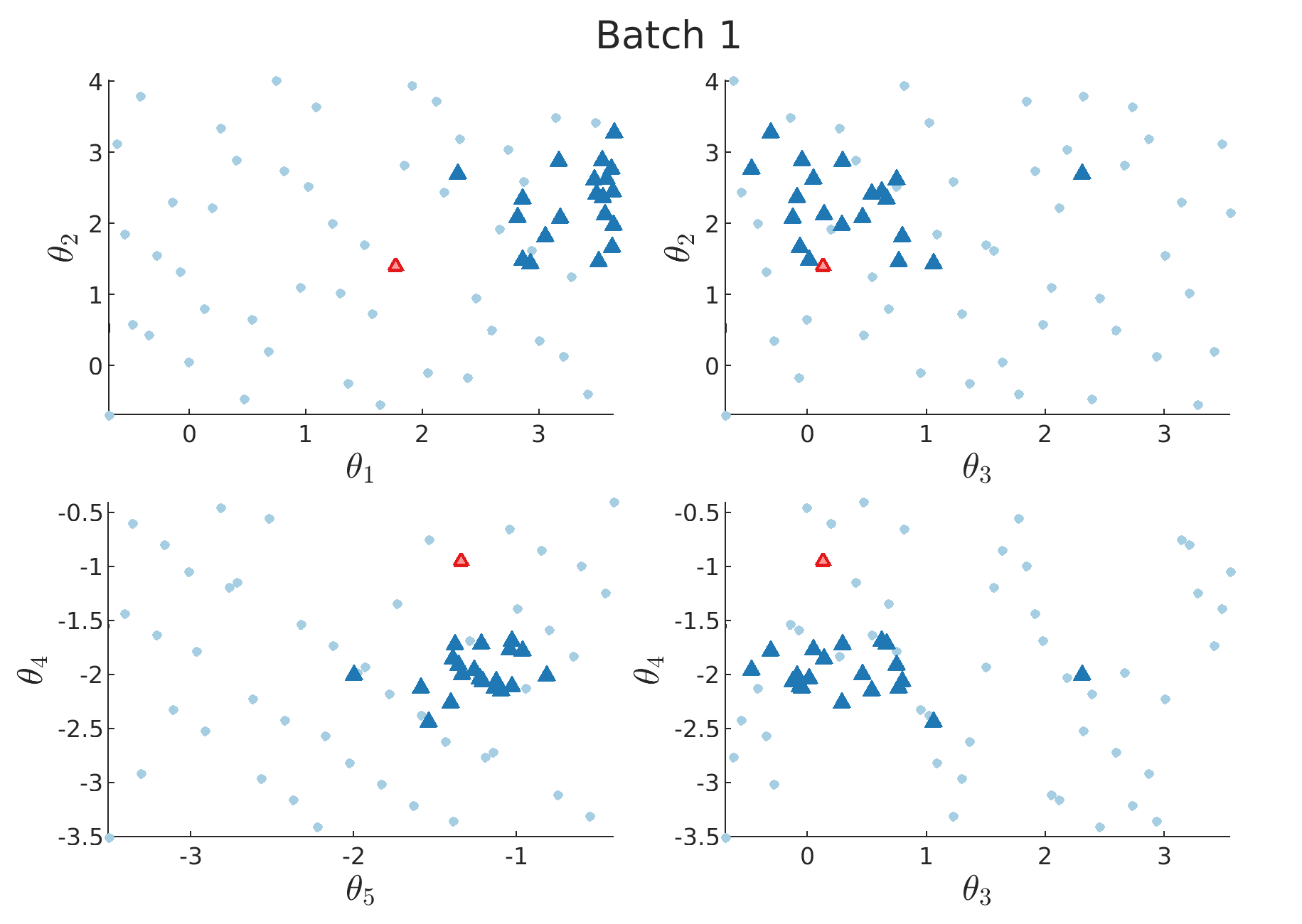}
  \caption{Iteration 1}
 \end{subfigure} 
 \begin{subfigure}[b]{0.495\textwidth}
  \centering 
  \includegraphics[trim={0 5px 0 20px},clip,width=\textwidth]{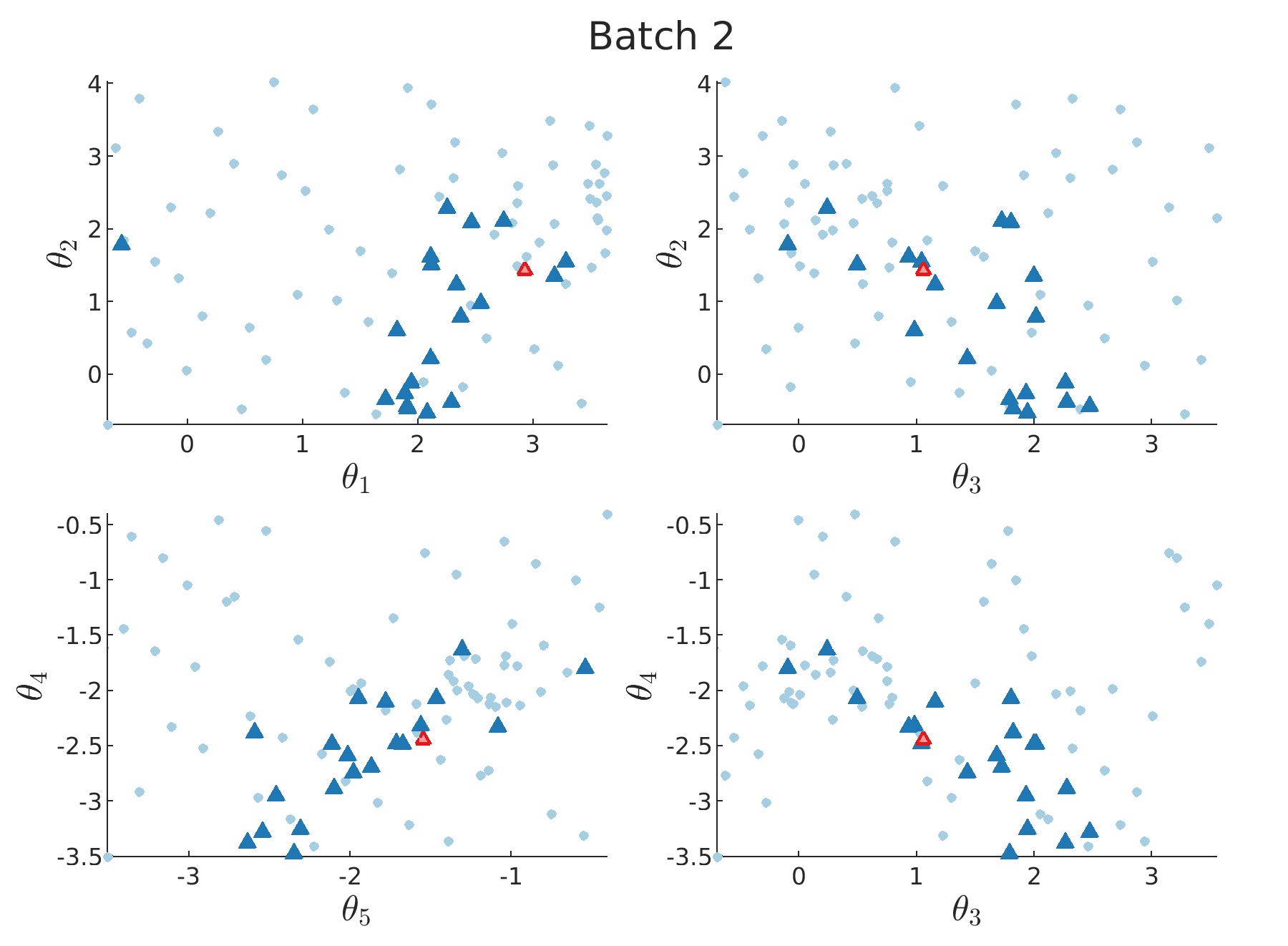}
  \caption{Iteration 2}
 \end{subfigure}
 \begin{subfigure}[b]{0.495\textwidth}
  \centering 
  \includegraphics[trim={0 5px 0 20px},clip,width=\textwidth]{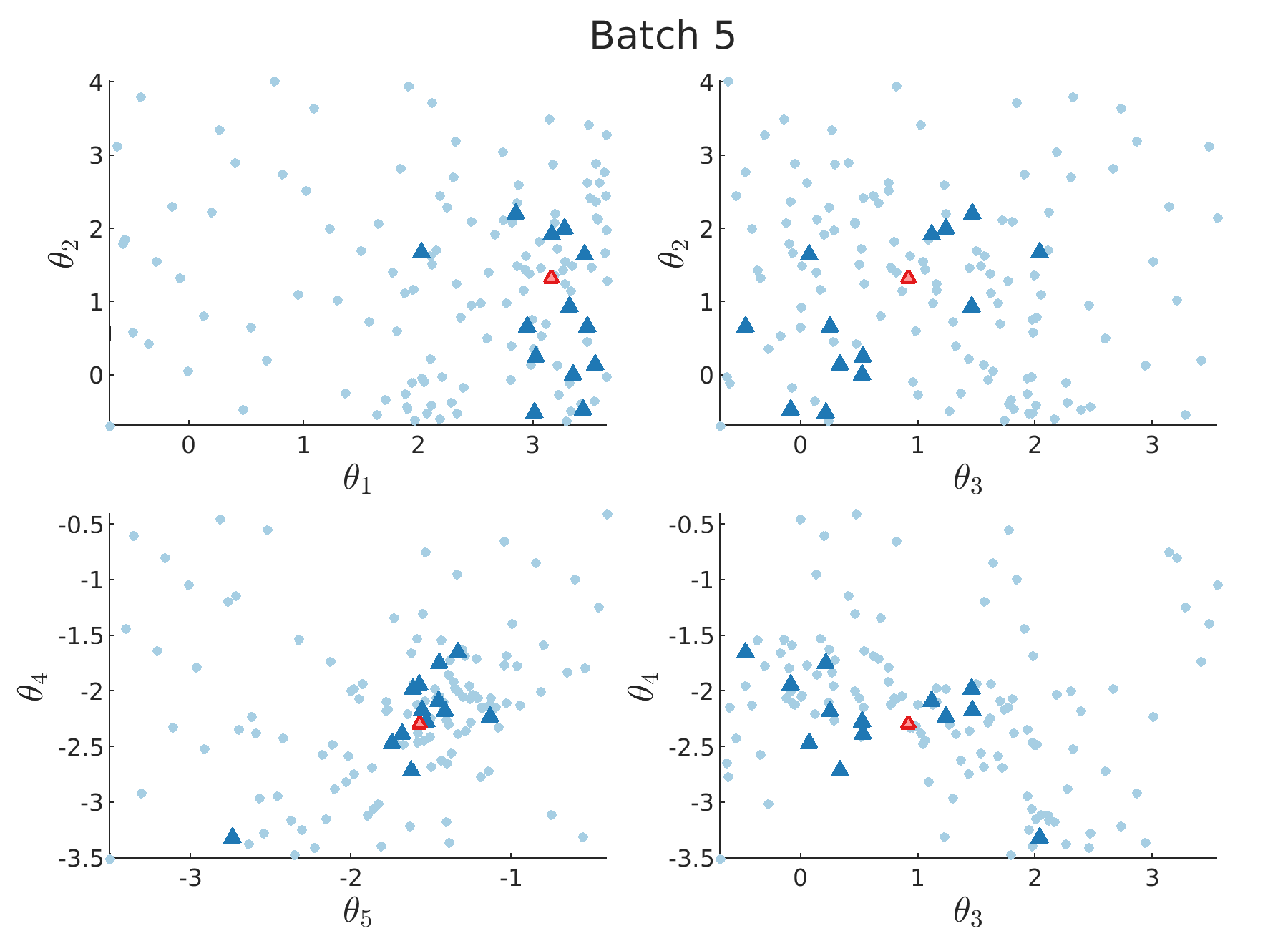}
  \caption{Iteration 5}
 \end{subfigure}
  \begin{subfigure}[b]{0.495\textwidth}
  \centering 
  \includegraphics[trim={0 5px 0 20px},clip,width=\textwidth]{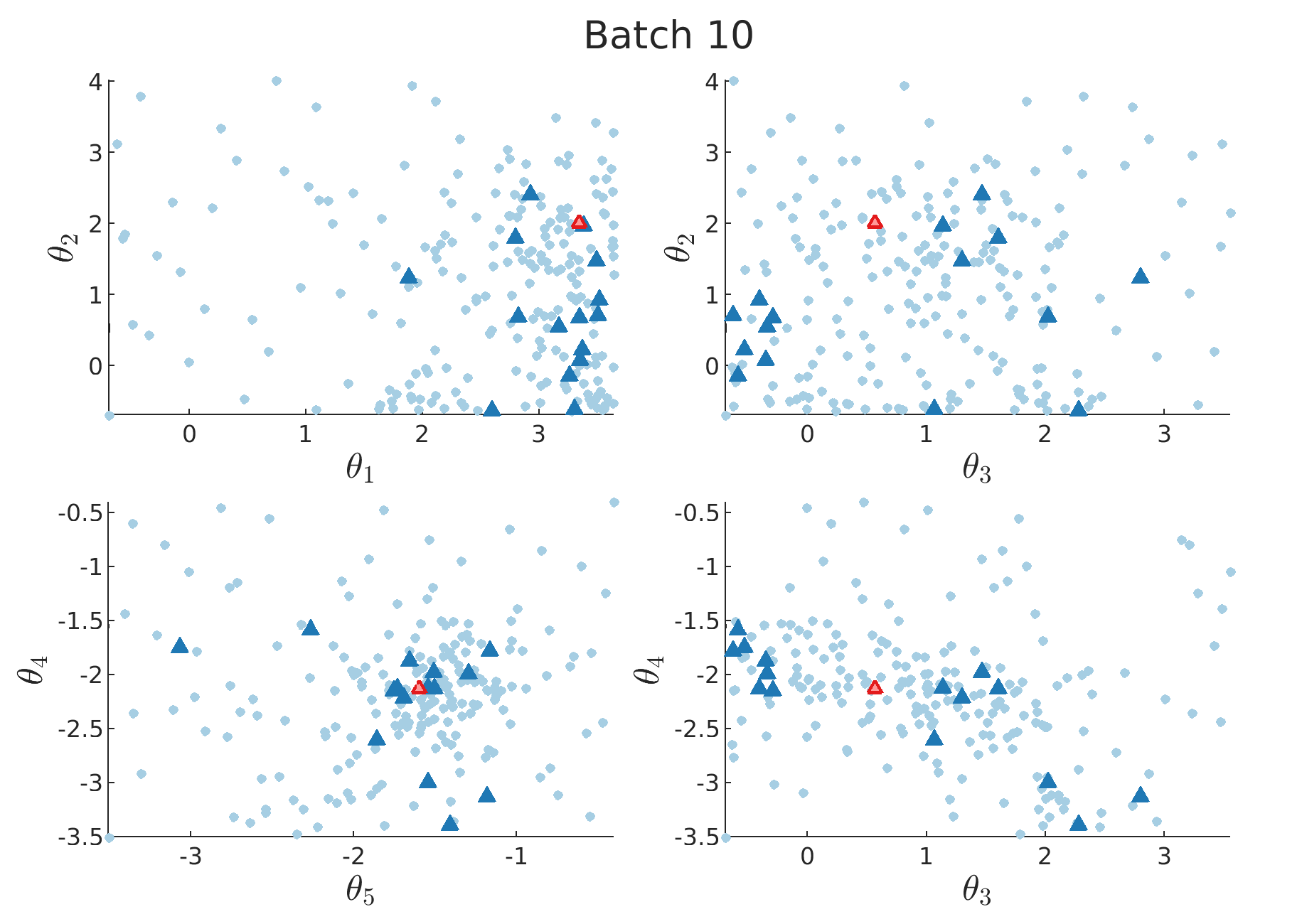}
  \caption{Iteration 10}
 \end{subfigure}
  \begin{subfigure}[b]{0.495\textwidth}
  \centering 
  \includegraphics[trim={0 5px 0 20px},clip,width=\textwidth]{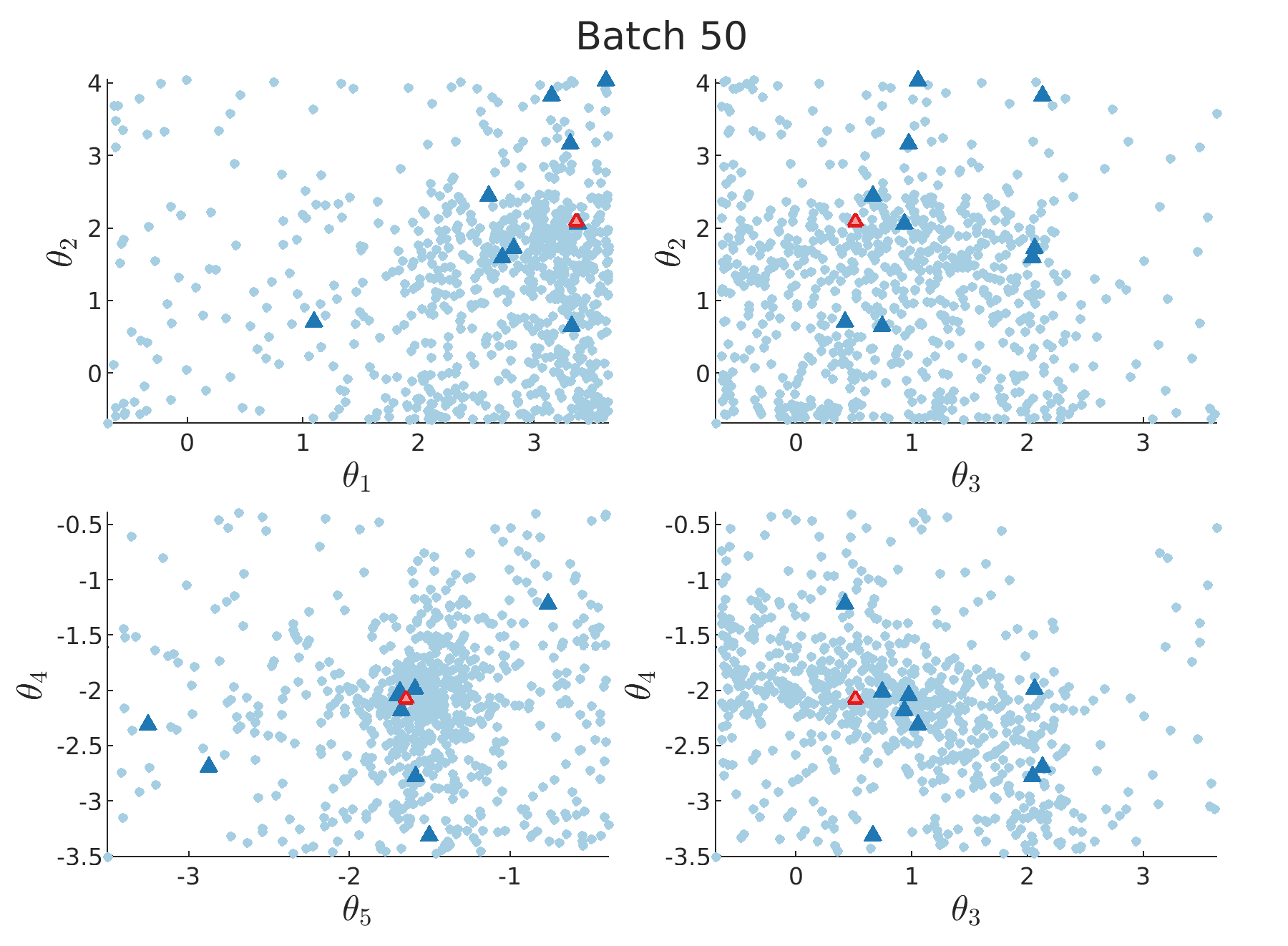}
  \caption{Iteration 50}
 \end{subfigure}
  \begin{subfigure}[b]{0.495\textwidth}
  \centering 
  \includegraphics[trim={0 5px 0 20px},clip,width=\textwidth]{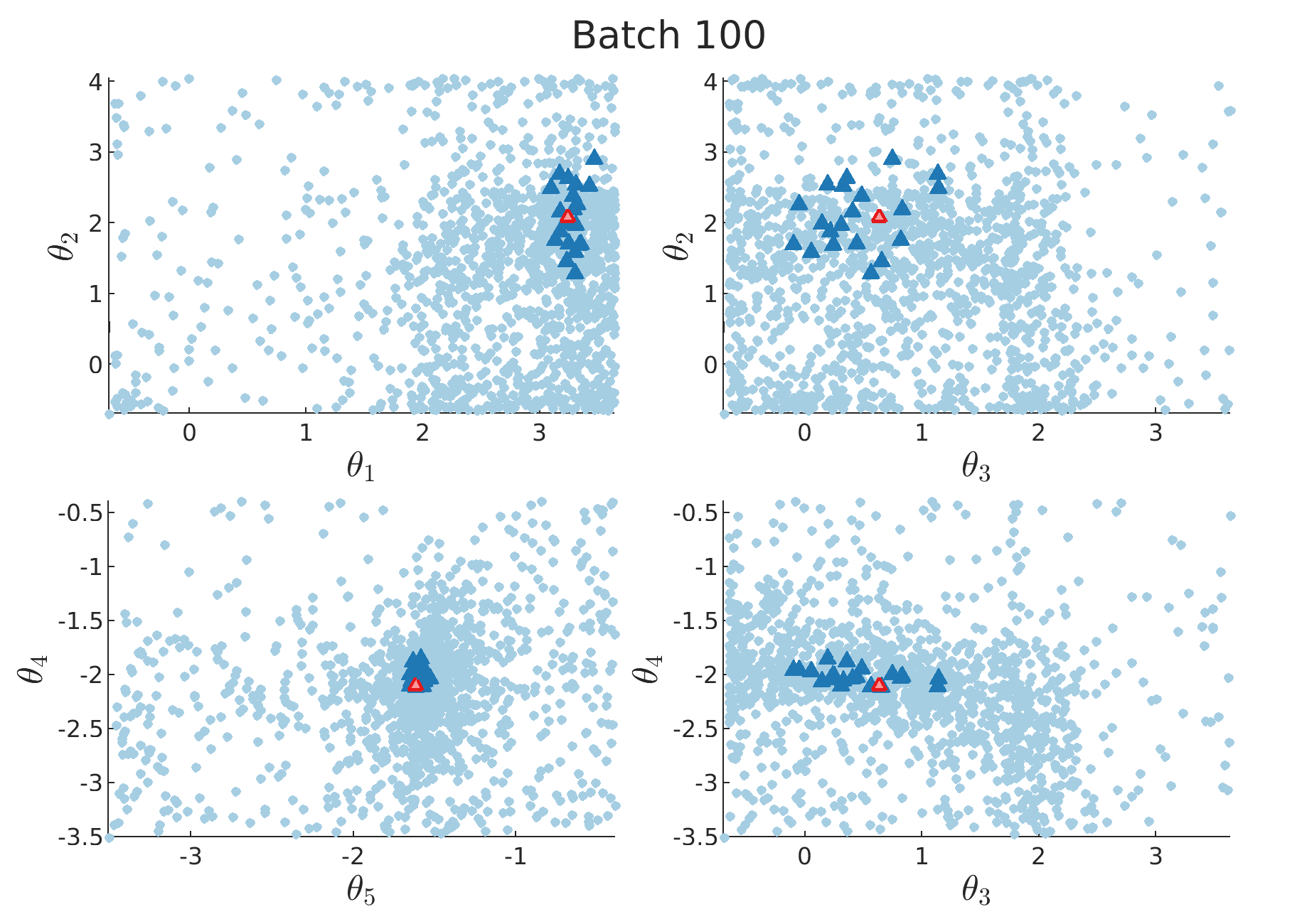}
  \caption{Iteration 100}
 \end{subfigure}     
 \caption{Scatter plots of candidate points chosen at various iterations of Bayes Optimization. Light blue points are candidates from earlier iterations and initial design, dark blue triangles are the candidates from current batch, and the red triangle is the previous mode.}
\label{fig:batchScatter}
\end{figure}

\begin{figure}
 \includegraphics[width=5in]{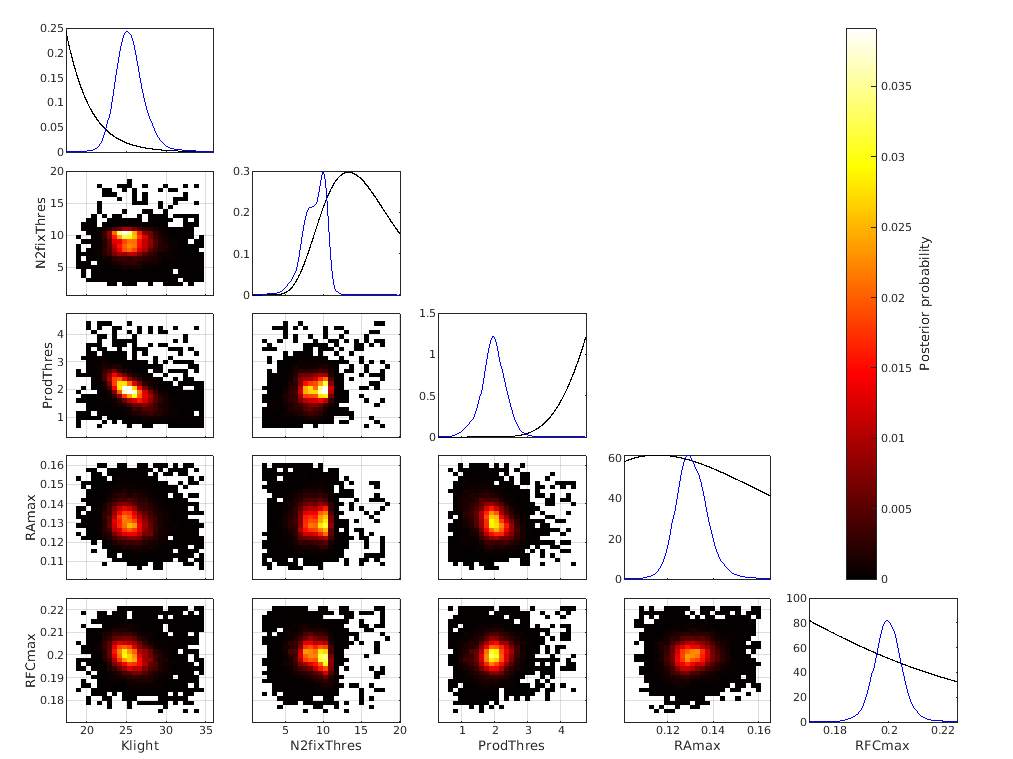}
 \caption{Parameter posterior marginal densities, including pairwise joint-marginal densities in the off-diagonal. On the diagonal, the blue line is the posterior density and the black line is the prior density scaled to match the posterior density. Joint probabilities are proportion of total samples in each bin, with total of $30^2=900$ bins.}
\label{fig:jointPosterior}
\end{figure}

\end{document}